\documentstyle[epsf,bibnorm]{lamuphys}
\makeatletter
\let\chapter\hid@chapter
\makeatother

\begin{document}
\pagenumbering{arabic}

\input{BoxedEPS.tex}

\SetRokickiEPSFSpecial
\HideDisplacementBoxes

\title{Electron-deuteron scattering in a relativistic
theory of hadrons}

\author{Daniel\,Phillips}

\institute{Department of Physics,\\
University of Maryland,\\
College Park, MD, 20742-4111}

\maketitle

\begin{abstract}
  We review a three-dimensional formalism that provides a systematic
  way to include relativistic effects including relativistic
  kinematics, the effects of negative-energy states, and the boosts of
  the two-body system in calculations of two-body bound-states. We
  then explain how to construct a conserved current within this
  relativistic three-dimensional approach. This general theoretical
  framework is specifically applied to electron-deuteron scattering
  both in impulse approximation and when the $\rho \pi \gamma$
  meson-exchange current is included. The experimentally-measured
  quantities $A$, $B$, and $T_{20}$ are calculated over the kinematic
  range that is probed in Jefferson Lab experiments. The role of both
  negative-energy states and meson retardation appears to be small in
  the region of interest.
\end{abstract}

\section{Introduction}

A number of the experiments being performed at the Thomas Jefferson
National Accelerator Facility (TJNAF) involve the elastic and
inelastic scattering of electrons off the deuteron at space-like
momentum transfers of the order of the nucleon mass. In building
theoretical models of these processes, relativistic kinematics and
dynamics would seem to be called for.  Much theoretical effort has
been spent constructing relativistic formalisms for the two-nucleon
bound state that are based on an effective quantum field theory
lagrangian.  If the usual hadronic degrees of freedom appear in the
lagrangian then this strategy is essentially a logical extension of
the standard nonrelativistic treatment of the two-nucleon system.

Furthermore, regardless of the momentum transfer involved, it is
crucial that a description of the deuteron be used which incorporates
the consequences of electromagnetic gauge invariance.  Minimally this
means that the electromagnetic current constructed for the deuteron
must be conserved.

Of course, the two-nucleon bound state can be calculated and a
corresponding conserved deuteron current constructed using
non-relativistic $NN$ potentials which are fit to the $NN$ scattering
data. This approach has met with considerable success.  (For some
examples of this program see Refs.~\cite{Ad93,Wi95}.) Our goal here is
to imitate such calculations---and, we hope, their success!---in a
relativistic framework. To do this we construct an $NN$ interaction,
place it in a relativistic scattering equation, and then fit the
parameters of our interaction to the $NN$ scattering data. We then
calculate the electromagnetic form factors of the deuteron predicted
by this $NN$ model. By proceeding in this way we hope to gain
understanding of the deuteron electromagnetic form factors in a model
in which relativistic effects, such as relativistic kinematics,
negative-energy states, boost effects, and relativistic pieces of the
electromagnetic current, are explicitly included at all stages of the
calculation.

This program could be pursued using a four-dimensional formalism based
on the Bethe-Salpeter equation.  Indeed, pioneering calculations of
electron-deuteron scattering using Bethe-Salpeter amplitudes were
performed by Zuilhof and Tjon almost twenty years
ago~\cite{ZT80,ZT81}. However, despite increases in computer power
since this early work the four-dimensional problem is still a
difficult one to solve. Since the $NN$ interaction is somewhat
phenomenological ultimately it is not clear that one gains greatly in
either dynamics or understanding by treating the problem
four-dimensionally.  Therefore, instead we will employ a
three-dimensional formalism that incorporates what we believe are the
important dynamical effects due to relativity at the momentum
transfers of interest.

We will use a three-dimensional (3D) formalism that, in principle, is
equivalent to the four-dimensional Bethe-Salpeter formalism. This
approach has been developed and applied in
Refs.~\cite{PW96,PW97,PW98}. In this paper we will focus on the
calculation of elastic electron-deuteron scattering. Here we review
the formalism for relativistic bound states and show how to construct
the corresponding electromagnetic current.  Calculations of elastic
electron-deuteron scattering are performed both in the impulse
approximation and with some meson-exchange currents included. The
results for the observables $A$, $B$ and $T_{20}$ are presented.

Many other 3D relativistic treatments of the deuteron dynamics which
are similar in spirit to that pursued here exist (see for instance
Refs.~\cite{vO95,Ar80}). Of these, our work is closest to that of
Hummel and Tjon~\cite{HT89,HT90,HT94}. However, in that work
approximations were employed for ingredients of the analysis, such as
the use of wave functions based on the 3D quasipotential propagator of
Blankenbecler-Sugar~\cite{BbS66} and Logunov-Tavkhelidze~\cite{LT63},
approximate boost operators, and an electromagnetic current which only
approximately satisfies current conservation. Calculations of elastic
electron-deuteron scattering also were performed by Devine and Wallace
using a similar approach to that pursued here~\cite{WD94}.  Here we
extend these previous analyses by use of our systematic 3D formalism.
In this way we can incorporate retardations into the interaction and
also use a deuteron electromagnetic current that is specifically
constructed to maintain the Ward-Takahashi identites.

The paper is organized as follows. In Section~\ref{sec-Section2} we
explain our reduction from four to three dimensions.  In
Section~\ref{sec-Section3} we present a four-dimensional equation
which is a modified version of the ladder Bethe-Salpeter equation.
This modified equation has the virtue that it, unlike the ladder BSE,
incorporates the correct one-body limit. By applying our
three-dimensional reduction technique to this four-dimensional
equation we produce an equation which has the correct one-body limit and
contains the correct physics of negative-energy states. In
Section~\ref{sec-Section4} we explain the various potentials that are
used in calculations of deuteron wave functions.  These can be divided
into two classes: instant potentials, and potentials that include
meson retardation.  Within either of these classes versions of the
potentials are constructed that do and do not include the effects of
negative-energy states, in order to display the role played by such
components of the deuteron wave function.  Section~\ref{sec-Section5}
discusses our 3D reduction of the electromagnetic current that
maintains current conservation. This completes the laying out of a
consistent formalism that includes the effects of relativity
systematically, has the correct one-body limits, and maintains current
conservation. In Section~\ref{sec-Section6} we apply this machinery to
the calculation of electron-deuteron scattering both in the impulse
approximation and when corrections due to some meson-exchange currents
are included.  Finally, discussion and conclusions are presented in
Section~\ref{sec-Section7}.

\section{The reduction to three dimensions}

\label{sec-Section2}

The Bethe-Salpeter equation, 
\begin{equation}
T=K + K G_0 T,
\label{eq:BSE}
\end{equation}
for the four-dimensional $NN$ amplitude $T$ provides a theoretical
description of the deuteron which incorporates relativity. 
Here $K$ is the Bethe-Salpeter kernel, and $G_0$ is the
free two-nucleon propagator.  In a strict quantum-field-theory
treatment, the kernel $K$ includes the infinite set of two-particle
irreducible $NN \rightarrow NN$ Feynman graphs.  

For the two-nucleon system an application of the full effective
quantum field theory of nucleons and mesons is impractical and
perhaps, since hadronic degrees of freedom are not fundamental,
inappropriate. In other words, the Bethe-Salpeter formalism may serve
as a theoretical framework within which some relativistic effective
interaction may be developed.  But, if the $NN$ interaction is only an
effective one, then it would seem to be equally appropriate to develop
the relativistic effective interaction within an equivalent
three-dimensional formalism which is obtained from the
four-dimensional Bethe-Salpeter formalism via some systematic
reduction technique.

One straightforward way to reduce the Bethe-Salpeter equation to three
dimensions is to approximate the kernel $K$ by an instantaneous
interaction $K_{\rm inst}$. For example, if $q=(q_0,\bf q)$ is the
relative four-momentum of the two nucleons then
\begin{equation}
K(q)=\frac{1}{q^2 - \mu^2} \qquad \rightarrow \qquad
K({\bf q})=-\frac{1}{{\bf q}^2 + \mu^2}.
\end{equation}
This, admittedly uncontrolled, approximation, yields from the Bethe-Salpeter
equation the Salpeter equation:
\begin{equation}
T_{\rm inst}=K_{\rm inst} + K_{\rm inst} \langle G_0 \rangle T_{\rm inst},
\label{eq:Salpeter}
\end{equation}
where the three-dimensional Salpeter propagator $\langle G_0 \rangle$ is
obtained by integrating over the time-component of relative momentum,
\begin{equation}
\langle G_0 \rangle=\int \frac{dp_0}{2 \pi} G_0(p;P).
\end{equation}
Throughout this paper we denote the integration over the zeroth
component of relative momenta, which is equivalent to consideration of
an equal-time Green's function, by angled brackets. We shall consider
only spin-half particles, and so
\begin{equation}
\langle G_0 \rangle=\frac{\Lambda_1^+ \Lambda_2^+}{E - \epsilon_1
- \epsilon_2} - \frac{\Lambda_1^- \Lambda_2^-}{E +  \epsilon_1
+ \epsilon_2};
\label{eq:aveG0}
\end{equation}
where $\Lambda^{\pm}$ are related to projection operators onto
positive and negative-energy states of the Dirac equation, $E$ is the
total energy, and $\epsilon_i=({\bf p}_i^2 + m_i^2)^{1/2}$. Note that
for spin-half particles, this propagator $\langle G_0 \rangle$ is not
invertible.

In order to systematize this kind of 3D reduction one must split the
4D kernel $K$ into two parts. One of these, $K_1$, is to be understood
as a three-dimensional interaction in the sense that it does not
depend on the zeroth component of relative four momentum~\footnote {Of
  course, this is not a covariant reduction, but covariance can be
  maintained by a suitable generalization of this idea~\cite{PW97}.}.
We then seek to choose this $K_1$ such that the 3D amplitude $T_1$
defined by
\begin{equation}
T_1=K_1 + K_1 \langle G_0 \rangle T_1,
\label{eq:3Dscatt}
\end{equation}
has the property that 
\begin{equation}
\langle G_0 \rangle T_1 \langle G_0 \rangle=\langle G_0 T G_0 \rangle.
\end{equation}

It is straightforward to demonstrate that such a $K_1$ is defined by
the coupled equations:
\begin{equation}
K_1 = \langle G_0 \rangle ^{-1} \langle G_0 K {\cal G} \rangle
\langle G_0 \rangle ^{-1} ,
\label{eq:K1}
\end{equation}
which is three-dimensional, and 
\begin{equation}
{\cal G} = G_0 + G_0 (K - K_1) {\cal G},
\label{eq:<G0>}
\end{equation}
which is four dimensional. The $K_1$ of Eq.~(\ref{eq:K1}) does this by
ensuring that
\begin{equation}
\langle {\cal G} \rangle=\langle G_0 \rangle.
\label{eq:calGeq}
\end{equation}
The formalism is systematic in the sense that, given a perturbative
expansion for the 4D kernel, $K$, a perturbative expansion for the 3D
kernel, $K_1$, can be developed.  At second order in the coupling this gives:
\begin{equation}
K_1^{(2)}=\langle G_0 \rangle^{-1} \langle G_0 K^{(2)} G_0 \rangle 
\langle G_0 \rangle^{-1}.
\label{eq:K12}
\end{equation}
In $++ \rightarrow ++$ states this is just the usual energy-dependent 
one-particle-exchange interaction of time-ordered perturbation theory,
but with relativistic kinematics, i.e. ignoring spin and isospin:
\begin{equation}
K_1^{(2)}=\frac{g^2}{2 \omega}\left[\frac{1}{E^+ - \epsilon_1 - 
\epsilon_2' - \omega} + (1 \leftrightarrow 2)\right],
\end{equation}
where $\omega$ is the on-shell energy of the exchanged particle.  Note
that $\langle G_0 \rangle$ must be invertible in order for the 3D
reduction to be consistent. (Similar connections between three and
four-dimensional approaches are discussed in
Refs.~\cite{LT63,Kl53,KL74,BK93B,LA97}.)

Equation~(\ref{eq:3Dscatt}) leads to an equation for the bound-state
vertex function:
\begin{equation}
\Gamma_1 = K_1 \langle G_0 \rangle \Gamma_1, 
\label{eq:3Deqn}
\end{equation}
where $\Gamma_1$ is the vertex function in the three-dimensional
theory. The 4D vertex function, $\Gamma$, and the
corresponding 3D one, $\Gamma_1$, are related via
\begin{equation}
G_0 \Gamma = {\cal G}\Gamma_1.
\label{eq:GcalGamma1}
\end{equation}

\section{The one-body limit}

\label{sec-Section3}

As mentioned above, and discussed many years ago by Klein~\cite{Kl53},
the propagator $\langle G_0 \rangle$ is not invertible and therefore
the above reduction is not consistent. We shall show in this section
that this difficulty is connected to the behavior of the
three-dimensional equation in the one-body limit.  In this limit we
allow one particle's mass to tend to infinity. We expect that the
amplitude $T_1$ then reduces to that given by the Dirac equation for a
light particle moving in the static field of the heavy particle. In
fact, this does not happen unless we include an infinite number of
graphs in the kernel of the integral equation Eq.~(\ref{eq:Salpeter}).

In fact, if a scattering equation with a kernel which contains only a
finite number of graphs is to possess the correct one-body limit, two
distinct criteria must be satisfied.  First the 3D propagator should
limit to the one-body propagator for one particle (the Dirac
propagator in this case) as the other particle's mass tends to
infinity.  Second, as either particle's mass tends to infinity, the
equation should become equivalent to one in which the interaction,
$K_1$, is static.

Equation~(\ref{eq:Salpeter})'s lack of either of these properties
stems from Eq.~(\ref{eq:BSE}) not having the correct one-body limit if
any kernel which does not include the infinite set of crossed-ladder
graphs is chosen~\cite{Gr82}. Solution of Eq.~(\ref{eq:BSE}) with such
a kernel is impractical in the $NN$ system. Nevertheless, the
contributions of crossed-ladder graphs to the kernel may be included
in an integral equation for $T$ by using a 4D integral equation for
$K$, the kernel of Eq.~(\ref{eq:BSE})
\begin{equation}
K = U + U G_C K. 
\label{eq:U} 
\end{equation}
Once $G_C$ is defined this equation defines a reduced kernel $U$ in
terms of the original kernel $K$.  The propagator $G_C$ is chosen so
as to separate the parts of the kernel $K$ that are necessary to
obtain the one-body limit from the parts that are not.  $U$ may then
be truncated at any desired order without losing the one-body limits.
The following 4D equation for the t-matrix is thus equivalent to
Eqs.~(\ref{eq:BSE}) and (\ref{eq:U}),
\begin{equation}
T = U + U (G_0 + G_C) T.
\label{eq:4DETampl}
\end{equation}  

We can now remedy the defects of our previous 3D reduction. Applying
the same 3D reduction used above to Eq.~(\ref{eq:4DETampl}) gives:
\begin{equation}
T_1 = U_1 + U_1 \langle G_0 + G_C \rangle T_1, 
\label{eq:3DETampl}
\end{equation}
where the 3D propagator is 
\begin{eqnarray}
  \langle G_0 + G_C \rangle&=&\frac{ \Lambda_1^+ \Lambda_2^+}{{P^0}^+
    - \epsilon_1 - \epsilon_2} - \frac{ \Lambda_1^+ \Lambda_2^-}{2
    \kappa_2^0 - {P^0}^+ + \epsilon_1 + \epsilon_2} \nonumber\\ 
&-&  \frac{ \Lambda_1^- \Lambda_2^+}{{P^0}^- - 2 \kappa_2^0 + \epsilon_1
    + \epsilon_2} - \frac{ \Lambda_1^- \Lambda_2^-}{{P^0}^- +
    \epsilon_1 + \epsilon_2},
\label{eq:aveG0GCgeneral}
\end{eqnarray}
and $\kappa_2^0$ is a parameter that enters through the construction
of $G_C$. This three-dimensional propagator was derived by Mandelzweig
and Wallace with the choice $\kappa_2^0 = P^0/2 - (m_1^2 -
m_2^2)/(2P^0)$~\cite{MW87,WM89}. With $\kappa_2^0$ chosen in this way
$\langle G_0 + G_C \rangle$ has the correct one-body limits as either
particle's mass tends to infinity and has an invertible form.  The
kernel $U_1$ is defined by Eqs.~(\ref{eq:K1}) and (\ref{eq:calGeq})
with the replacements $G_0 \rightarrow G_0 + G_C$, $K \rightarrow U$,
and $K_1 \rightarrow U_1$.

Here we are interested in the scattering of particles of equal mass
and so we make a different choice for $\kappa_2^0$. Specifically,
\begin{equation}
\kappa_2^0 = \frac{P^0 - \epsilon_1 + \epsilon_2}{2}.
\label{eq:kappachoice}
\end{equation}
This form avoids the appearance of unphysical singularities when
electron-deuteron scattering is calculated~\cite{PW98}. It yields a
two-body propagator:
\begin{equation}
  \langle G_0 + G_C \rangle=\frac{ \Lambda_1^+ \Lambda_2^+}{{P^0}^+
    - \epsilon_1 - \epsilon_2} - \frac{ \Lambda_1^+ \Lambda_2^-}{2
    \epsilon_2} - \frac{ \Lambda_1^- \Lambda_2^+}{2 \epsilon_1} -
  \frac{ \Lambda_1^- \Lambda_2^-}{{P^0}^- + \epsilon_1 + \epsilon_2},
\label{eq:aveG0GC}
\end{equation}
which is consistent with that required by low-energy theorems for
Dirac particles in scalar and vector fields~\cite{Ph97}. Another way
of saying this is to realize that if we compare the the $++
\rightarrow ++$ piece of the amplitude
\begin{equation}
V_1 \langle G_0 + G_C \rangle V_1
\end{equation}
to the amplitude obtained at fourth order in the full 4D field theory
then the contribution of negative-energy states agrees at leading
order in $1/M$~\cite{PW98}. 

\begin{figure}[h]
\centerline{\BoxedEPSF{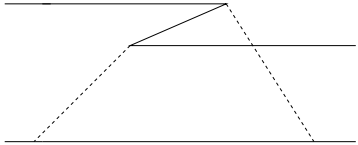 scaled 350}}

\caption{One example of a Z-graph which is included in our
3D equation (\ref{eq:3DET}).}
\label{fig-Zgraph}  
\end{figure}

For bound states the argument of the previous section leads to the 3D
equation:
\begin{equation}
\Gamma_1=U_1 \langle G_0 + G_C \rangle \Gamma_1.
\label{eq:3DET}
\end{equation}
Equation (\ref{eq:3DET}) is a bound-state equation which incorporates
relativistic effects and the physics of negative-energy states.
For instance, fig.~\ref{fig-Zgraph} is one example of a
graph which is included if Eq.~(\ref{eq:3DET}), even if only the
lowest-order kernel $U_1^{(2)}$ is used, because of our careful treatment
of the one-body limit.

\section{Results for the deuteron}

\label{sec-Section4}

To calculate observables in the deuteron we now consider two types of
kernels $U_1$, both of which are calculated within the framework
of a one-boson exchange model for the $NN$ interaction:
\begin{enumerate}
\item $U_1=U_{\rm inst}$, the instantaneous interaction.

\item A kernel $U_1^{(2)}$ which is a retarded interaction.  This is
  obtained from Eq.~(\ref{eq:K12}) by the substitutions $K_1^{(2)}
  \rightarrow U_1^{(2)}$ and $G_0 \rightarrow G_0 + G_C$.
\end{enumerate}
These interactions are used in a two-body equation with the full ET
Green's function given by Eq.~(\ref{eq:3DET}), and also in an equation
in which only the $++$ sector is retained. For the instant
interaction, we follow the practice of Devine and Wallace~\cite{WD94}
and switch off couplings between the $++$ and $--$ sectors, and
between the $+-$ and $-+$ sectors.  A partial justification of this
rule follows from an analysis of the static limit of our 3D retarded
interaction.

The mesons in our one-boson exchange model are the $\pi(138)$, the
$\sigma(550)$, the $\eta(549)$, the $\rho(769)$, the $\omega(782)$,
and the $\delta(983)$.  All the parameters of the model, except for
the $\sigma $ coupling, are taken directly from the Bonn-B fit to the
$NN$ phase shifts~\cite{Ma89}---which is a fit performed using a
relativistic wave equation and relativistic propagators for the
mesons. The $\sigma$ coupling is varied so as to achieve the correct
deuteron binding energy for each interaction considered. Of course, we
should refit the parameters of our $NN$ interaction using our
different scattering equations. However, for a first estimate of the
importance of negative-energy states and retardation we adopt this
simpler approach to constructing the interaction. Work on improving
the $NN$ interaction model is in progress~\cite{PW98B}.

Once a particular interaction is chosen, the integral equation
(\ref{eq:3DET}) is solved for the bound-state energy.  In each
calculation, the $\sigma$ coupling is adjusted to get the correct
deuteron binding energy, producing the results (accurate to three
significant figures) given in Table~\ref{table-sigmacoupling}.  The
value given for the instant calculation with positive-energy states
alone is that found in the original Bonn-B fit. In all other cases the
$\sigma$ coupling must be adjusted to compensate for the inclusion of
retardation, the effects of negative-energy states, etc.  We believe
that this adjustment of the scalar coupling strength is sufficient to
get a reasonable deuteron wave function. The static properties of this
deuteron are very similar to those of a deuteron calculated with the
usual Bonn-B interaction.

With the bound-state wave function in the center-of-mass frame has
been determined in this fashion, it is a simple matter to solve the
integral equation (\ref{eq:3DET}) in any other frame.  We choose to
calculate electron-deuteron scattering in the Breit frame. The
interaction is recalculated in the Breit frame for a given $Q^2$, and
then the integral equation is solved with this new interaction.
Because the formalism we use for reducing the four-dimensional
integral equation to three dimensions is {\it not} Lorentz invariant
there is a violation of Lorentz invariance in this calculation.
Estimations of the degree to which Lorentz invariance is violated are
displayed in Ref.~\cite{PW98}.

\begin{table}[htbp]
\caption{Sigma coupling required to produce the correct deuteron binding
energy in the four different models under consideration here.}
\label{table-sigmacoupling}
\begin{center}
\begin{tabular}{|c|c|c|}
\hline
{\bf Interaction}   &  {\bf States included} & {\bf $g_\sigma^2/4 \pi$}  \\ 
\hline
Instant      &       ++         &     8.08            \\ \hline
Retarded     &       ++         &     8.39            \\ \hline
Instant      &       All        &     8.55            \\ \hline
Retarded     &       All        &     8.44            \\  \hline
\end{tabular}
\end{center}
\end{table}

\section{Current conservation}

\label{sec-Section5}

\subsection{Currents in the three-dimensional formalism}

\label{sec-3Dcurrents}

As discussed in the Introduction, we now want to compare the predictions
of this formalism with experimental data gained in electron scattering
experiments. In calculating the interaction of the electron
with the hadronic bound state it is crucial to derive a 
3D reduction of the electromagnetic current which is consistent with
the reduction of the scattering equation we have chosen to use here.

The current in the full four-dimensional formalism is obtained by
coupling photons everywhere on the right-hand side of
Eq.~(\ref{eq:BSE}). This produces the following gauge-invariant
result for the photon's interaction with the bound state:
\begin{eqnarray}
{\cal A}_\mu&=&\bar{\Gamma}(P') G_0(P') J_\mu G_0 (P) \Gamma(P) \nonumber\\
&+& \bar{\Gamma}(P') G_0(P') K_\mu^\gamma G_0 (P) \Gamma(P),
\label{eq:gi4damp}
\end{eqnarray}
where $P$ and $P'$ are the initial and final total four-momenta of the
deuteron bound state.  Here $J_\mu$ contains the usual one-body
currents and $K_\mu^\gamma$ represents two-body contributions which
are necessary for maintaining the Ward-Takahashi identities. All
integrals implicitly are four-dimensional. The connection to the
three-dimensional amplitude, $\Gamma_1$, obtained from
Eq.~(\ref{eq:3DET}) is made by inserting Eq.~(\ref{eq:GcalGamma1})
into Eq.~(\ref{eq:gi4damp}), giving
\begin{equation}
{\cal A}_\mu=\bar{\Gamma}_1(P') \langle {\cal G}(P') 
\left[J_\mu + K_\mu^\gamma \right] {\cal G}(P) \rangle \Gamma_1(P).
\label{eq:Amu}
\end{equation}
Once the effective operator $\langle {\cal G}(P') \left[J_\mu +
K_\mu^\gamma \right] {\cal G}(P) \rangle$ is calculated the
expression (\ref{eq:Amu}) involves only three-dimensional integrals.

Since ${\cal G}$ is an infinite series in $K-K_1$ this result would
not be much help on its own. But, given a result for $\Gamma_1$
obtained by systematic expansion of $K_1$, the amplitude ${\cal
  A}_\mu$ can be analogously expanded in a way that maintains current
conservation. $K_1$ as defined by Eq.~(\ref{eq:K1}) is an infinite
series and the condition (\ref{eq:calGeq}) is imposed order-by-order
in the expansion in $K-K_1$ defines $K_1$ to some finite order. The
question is: Does a corresponding 3D approximation for the current
matrix element (\ref{eq:Amu}) exist that maintains the Ward-Takahashi
identities of the theory?  {\it It turns out that the current matrix
  element (\ref{eq:Amu}) is conserved if ${\cal G} (J_\mu +
  K^\gamma_\mu) {\cal G}$ on the right-hand side of Eq.~(\ref{eq:Amu})
  is expanded to a given order in the coupling constant and the kernel
  $K_1$ used to define $\Gamma_1$ is obtained from
  Eq.~(\ref{eq:calGeq}) by truncation at the same order in the
  coupling constant.}

This is done by splitting the right-hand side of Eq.~(\ref{eq:Amu})
into two pieces, one due to the one-body current $J_\mu$, and one due
to the two-body current $K^\gamma_\mu$. If $K_1$ has been truncated at
lowest order---i.e., $K_1=K_1^{(2)}$---then, in the $J_\mu$ piece, we
expand the $\cal G$s and retain terms up to the same order in
$K^{(2)}-K_1^{(2)}$.  A piece from the two-body current, in which we
write ${\cal G}=G_0$, is added to this. That is, we define our
second-order approximation to ${\cal A}_\mu$, ${\cal A}_\mu^{(2)}$, by
\begin{eqnarray}
{\cal A}^{(2)}_\mu&=&\bar{\Gamma}_1(P') \langle G^\gamma_{0 \mu}
\rangle \Gamma_1(P) \nonumber \\ &+& \bar{\Gamma}_1(P') \langle G_0(P') (K^{(2)}(P')-K^{(2)}_1(P'))
G^\gamma_{0 \mu} \rangle \Gamma_1(P)
\nonumber\\
&+& \bar{\Gamma}_1(P') \langle G^{\gamma}_{0 \mu} (K^{(2)}(P)-K^{(2)}_1(P)) G_0(P) \rangle
 \Gamma_1(P)
\nonumber\\ 
&+& \bar{\Gamma}_1(P') \langle G_0(P') K^{\gamma (2)}_\mu G_0(P) \rangle
\Gamma_1(P),
\label{eq:A2mu}
\end{eqnarray}
where ${G_0^\gamma}_\mu=G_0(P') J_\mu G_0(P)$.  It can now be shown
that if Eq.~(\ref{eq:calGeq}) expanded to second order defines
$K_1^{(2)}$, the corresponding amplitude for electromagnetic
interactions of the bound state, as defined by Eq.~(\ref{eq:A2mu}),
exactly obeys
\begin{equation}
Q^\mu {\cal A}^{(2)}_\mu=0.
\label{eq:A1WTI}
\end{equation}

It is straightforward to check that the same result holds if
Eq.~(\ref{eq:calGeq}) for $K_1$ is truncated at fourth order, while
the one-body and two-body current pieces are expanded to fourth order.

\begin{figure}[h]
\centerline{\BoxedEPSF{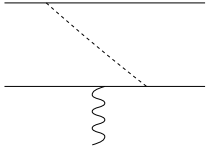 scaled 350}}

\caption{One example of a two-body current that is required in our formalism
in order to maintain current conservation.}
\label{fig-inflight}  
\end{figure}

The amplitude ${\cal A}_\mu^{(2)}$ includes contributions from
diagrams where the photon couples to particles one and two while
exchanged quanta are ``in-flight''. These contributions are of two
kinds.  Firstly, if the four-dimensional kernel $K$ is dependent on
the total momentum, or if it involves the exchange of charged
particles, then the WTIs in the 4D theory require that $K_\mu^\gamma$
contain terms involving the coupling of the photon to internal lines
in $K$.  Secondly, even if such terms are not present, terms arise in
the three-dimensional formalism where the photon couples to particles
one and two while an exchanged meson is ``in-flight''.  These must be
included if our 3D approach is to lead to a conserved current. (See
Fig.~\ref{fig-inflight} for one such mechanism.)

A special case of the above results occurs when retardation effects
are omitted, i.e., the kernel $K_1=K_{\rm inst}$, is chosen, and the
bound-state equation (\ref{eq:3Deqn}) is solved to get the vertex
function $\Gamma_1=\Gamma_{\rm inst}$. Then a simple conserved current
is found:
\begin{equation}
{\cal A}_{{\rm inst},\mu}=\bar{\Gamma}_{\rm inst}(P') \langle
G_{0 \mu}^\gamma \rangle \Gamma_{\rm inst}(P) + \bar{\Gamma}_{\rm
inst}(P') \langle G_0(P') \rangle {K^\gamma_{\rm inst}}_\mu 
\langle G_0(P) \rangle \Gamma_{\rm inst}(P),
\label{eq:instantme}
\end{equation}
where we have also replaced the meson-exchange current kernel 
$K^\gamma_\mu$ by the instant approximation to it.

\subsection{Current conservation in the 4D formalism with $G_C$}

In Ref.~\cite{PW98} we showed how to construct a conserved current
consistent with the 4D equation 
\begin{equation}
\Gamma=U (G_0 + G_C) \Gamma.
\label{eq:4DET}
\end{equation}
This turns out to be a moderately complicated exercise, because the
propagator $G_C$ depends on the three-momenta of particles one and
two, not only in the usual way, but also through the choice
(\ref{eq:kappachoice}) made for $\kappa_2^0$ above. However, a 4D
current ${\cal G}_{0,\mu}^\gamma = {G_0^\gamma}_\mu +
{G_C^\gamma}_\mu$ corresponding to the free Green's function $G_0 +
G_C$ can be constructed. Its form is displayed in Ref.~\cite{PW98} and
is not really germane to our purposes here, for, as we shall see
hereafter, only certain pieces of the current ${\cal
  G}_{0,\mu}^\gamma$ are actually used in our calculations.

\subsection{Reduction to 3D and the ET current} 

Having constructed a 4D current for the formalism involving $G_C$ that
obeys the required Ward-Takahashi identity, we can apply the reduction
formalism of Section~\ref{sec-3Dcurrents} to obtain the currents
corresponding to the 3D reduction of this 4D theory.  The result is:
\begin{eqnarray} 
{\cal A}^{(2)}_\mu&=&\bar{\Gamma}_{1,{\rm ET}}(P') \langle {\cal
G}^\gamma_{0,\mu} \rangle \Gamma_{1,{\rm ET}}(P) \nonumber\\
&+& \bar{\Gamma}_{1,{\rm ET}}(P') \langle
(G_0 + G_C)(P') (K^{(2)}(P')-U_1^{(2)}(P')) {\cal G}^\gamma_{0,\mu} 
\rangle \Gamma_{1,{\rm ET}}(P)
\nonumber\\
&+& \bar{\Gamma}_{1,{\rm ET}}(P') \langle {\cal G}^{\gamma}_{0,\mu} 
(K^{(2)}(P)-U_1^{(2)}(P)) (G_0 + G_C)(P) \rangle \Gamma_{1,{\rm ET}}(P)
\nonumber\\ 
&+& \bar{\Gamma}_{1,{\rm ET}}(P') \langle (G_0 + G_C)(P') K^{\gamma (2)}_\mu 
(G_0 + G_C)(P) \rangle \Gamma_{1,{\rm ET}}(P),
\label{eq:general3DETcurrent(2)}
\end{eqnarray}
where $\Gamma_{1,{\rm ET}}$ is the solution of Eq.~(\ref{eq:3DET}) with
$U_1=U_1^{(2)}$.
This current obeys the appropriate Ward-Takahashi identity.  In fact in
one-boson exchange models the only contributions to $K^{\gamma
  (2)}_\mu$ give rise to isovector structures, and so their
contribution to electromagnetic scattering off the deuteron is zero.

\subsection{Impulse-approximation current based on the instant
approximation to ET formalism}

Just as in the case of the Bethe-Salpeter equation, if the instant
approximation is used to obtain a bound-state equation with an instant
interaction from Eq.~(\ref{eq:4DET}) then a corresponding simple
conserved impulse current can be constructed:
\begin{equation}
{\cal A}_{{\rm inst},\mu}=\bar{\Gamma}_{\rm inst} 
\langle {\cal G}^\gamma_{0,\mu} \rangle
\Gamma_{\rm inst}.
\end{equation}

Now we note that the full result for ${\cal G}^\gamma_{0,\mu}$ was
constructed in order to obey Ward-Takahashi identities in the full
four-dimensional theory. It is not necessary to use this result if we
are only concerned with maintaining WTIs at the three-dimensional
level in the instant approximation.  Therefore we may construct the
corresponding current
\begin{eqnarray}
&& {\cal G}_{{\rm inst},\mu}^\gamma({\bf p}_1,{\bf p}_2;P,Q)=i \langle d_1(p_1)
d_2(p_2+Q) j^{(2)}_\mu d_2(p_2) \nonumber\\
&& \qquad + d_1(p_1) d_2^{\tilde{c}}(p_2+Q) j^{(2)}_{c,\mu} d_2^c(p_2) \rangle
+ (1 \leftrightarrow 2).
\end{eqnarray}
Here $d_i$ is the Dirac propagator for particle $i$, and $j_\mu=q
\gamma_\mu$ is the usual one-body current, with $q$ is the charge of
the particle in question. Meanwhile $d_i^c$ is a one-body Dirac
propagator used in $G_C(P)$ to construct the approximation to the
crossed-ladder graphs.  Correspondingly, $d_i^{\tilde{c}}$ appears in
$G_C(P+Q)$, which does {\it not} equal $d_i^c$, even if particle $i$
is not the nucleon struck by the photon.  Finally,
\begin{equation}
j^{(2)}_{c,\mu}= q_2 \gamma_\mu - \tilde{j}^{(2)}_\mu,
\end{equation}
where
\begin{equation}
  \tilde{j}^{(2)}_\mu=q_2 \frac{\hat{p}_{2 \mu}' + \hat{p}_{2
      \mu}}{\epsilon_2' + \epsilon_2} {\gamma_{2}}_0,
\label{eq:tildej2}
\end{equation}
with $\hat{p}_2=(\epsilon({\bf p}_2),{\bf p}_2)$.  (For further
explanation of these quantities and the necessity of their appearance
here the reader is referred to Ref.~\cite{PW98}.)

If a vertex function $\Gamma_{{\rm inst}}$ is constructed to be a
solution to Eq.~(\ref{eq:3DET}) with an instant interaction then the
three-dimensional hadronic current:
\begin{equation}
{\cal A}_{{\rm inst},\mu}= \bar{\Gamma}_{\rm inst} {\cal G}_{{\rm inst},\mu}^\gamma \Gamma_{\rm inst}
\label{eq:instAmu}
\end{equation}
is conserved.  This current is simpler than the full ET current and
omits only effects stemming from retardation in the current. Our
present calculations are designed to provide an assessment of the role
of negative-energy states and retardation effects in the vertex
functions. Therefore we use the simple current (\ref{eq:instAmu}) in
{\it all} of our calculations here---even the ones where $\Gamma_1$ is
calculated using a retarded two-body interaction. The effects stemming
from retardation in the current are expected to be minor, and so we
expect this to be a good approximation to the full current in the
three-dimensional theory. Future calculations should be performed to
check the role of meson retardation in that current.

\section{Results for electron-deuteron scattering}

\label{sec-Section6}

\subsection{Impulse approximation}

We are now ready to calculate the experimentally observed deuteron
electromagnetic form factors $A$ and $B$, and the tensor polarization
$T_{20}$. These are straightforwardly related to the charge,
quadrupole, and magnetic form factors of the deuteron, $F_C$, $F_Q$,
and $F_M$.  These form factors in turn are related to the Breit frame
matrix elements of the current ${\cal A}_\mu$ discussed in the
previous section,
\begin{eqnarray}
F_C&=&\frac{1}{3\sqrt{1 + \eta}e} (\langle 0|{\cal A}^0| 0 \rangle 
+ 2 \langle +1|{\cal A}^0|+1 \rangle),\\
F_Q&=&\frac{1}{2 \eta \sqrt{1 + \eta} e} (\langle 0|{\cal A}^0| 0 \rangle
- \langle +1|{\cal A}^0|+1 \rangle),\\
F_M&=& \frac{-1}{\sqrt{2 \eta (1 + \eta)}e} \langle +1|{\cal A}_+|0\rangle,
\end{eqnarray}
where $|+1 \rangle$, $|0 \rangle$ and $|-1 \rangle$ are the 
three different spin states of the deuteron.

We take the wave functions constructed for the four different
interactions of Section~\ref{sec-Section4} and insert them into the
expression (\ref{eq:instAmu}). In using any of the interactions
obtained with only positive-energy state propagation we drop all
pieces of the operator ${\cal G}^\gamma_{{\rm inst},\mu}$ in
negative-energy sectors.

The single-nucleon current used in these calculations is the usual one
for extended nucleons.  We choose to parametrize the single-nucleon
form factors $F_1$ and $F_2$ via the 1976 Hohler fits~\cite{Ho76}.
Choosing different single-nucleon form factors does not affect our
qualitative conclusions, although it has some impact on our
quantitative results for $A$, $B$, and $T_{20}$.

Using this one-body current we then calculate the current matrix
elements via Eq.~(\ref{eq:Amu}). This is a conserved current if the
vertex function $\Gamma_1$ is calculated from an instant potential.
However, if a potential including meson retardation is used it
violates the Ward-Takahashi identities by omission of pieces that are
required because of the inclusion of retardation effects in the
calculation. Work is in progress to estimate the size of these
effects.

\begin{figure}[htbp]
\centerline{\BoxedEPSF{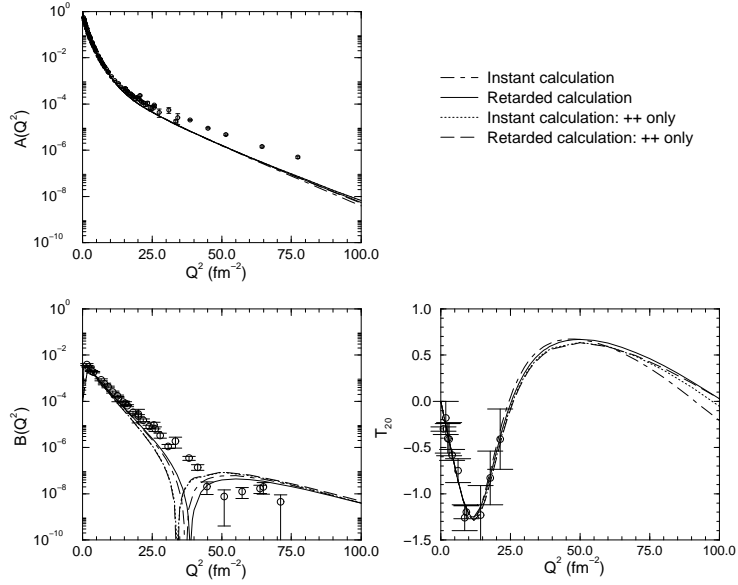 scaled 500}}
\caption{The form factors $A(Q^2)$ and $B(Q^2)$ and the tensor
  polarization $T_{20}$ for the deuteron calculated in impulse
  approximation. The dash-dotted line represents a calculation using a
  vertex function generated using the instant interaction. Meanwhile
  the solid line is the result obtained with the retarded vertex
  function. The dotted and long dashed lines are obtained by
  performing a calculations with instant and retarded interactions in which
  no negative-energy states are included.}
\label{fig-IA}
\end{figure}

The results for the impulse approximation calculation of the
experimental observables $A$, $B$, and $T_{20}$ are displayed in
Fig.~\ref{fig-IA}. We also show experimental data from
Refs.~\cite{El69,Ar75,Si81,Cr85,Pl90} for $A$, from
Refs.~\cite{Si81,Cr85,Au85,Bo90} for $B$ and from
Ref.~\cite{Sc84} for $T_{20}$.  A number of
two-body effects must be added to our calculations before they can be
reliably compared to experimental data. However, even here we see the
close similarity of the results for these observables in all four
calculations.  The only really noticeable difference occurs at the
minimum in $B$. There, including the negative-energy states in the
calculation shifts the minimum to somewhat larger $Q^2$. A similar
effect was observed by van Orden {\it et al.}~\cite{vO95} in
calculations of electron-deuteron scattering using the spectator
formalism. However, note that here, in contradistinction to the
results of Ref.~\cite{vO95}, the inclusion of negative-energy states
does {\it not} bring the impulse approximation calculation into
agreement with the data.

The fact that negative-energy states seem to have a smaller effect on
observables in the ET analysis than in the spectator analysis of van
Orden et al.~\cite{vO95} is somewhat surprising since our ``ET''
propagator has twice the negative-energy state propagation amplitude
of the spectator propagator. Thus, other differences between the ET and
spectator models, not just differences in the role of negative-energy
states in the two approaches, appear to be responsible for Ref.~\cite{vO95}'s
success in reproducing the minimum in $B$.

For the tensor polarization $T_{20}$ the different models produce
results which are very similar. This suggests that this observable is
fairly insensitive to dynamical details of the deuteron model, at
least up to $Q^2=4 \,\, \rm{GeV}^2$.

\subsection{Meson-exchange currents}

As $Q^2$ increases the cross-section due to the impulse approximation
diagrams drops precipitously. Thus we expect that in some regime other
interactions may become competitive with the impulse mechanism.  One
such possibility is that the photon will couple to a meson while that
meson is in flight.  Because of the deuteron's isoscalar nature and
the conservation of G-parity, the lowest mass state which can
contribute in such meson-exchange current (MEC) diagrams is one where
the photon induces a transition from a $\pi$ to a $\rho$.

\begin{figure}[htbp]
\centerline{\BoxedEPSF{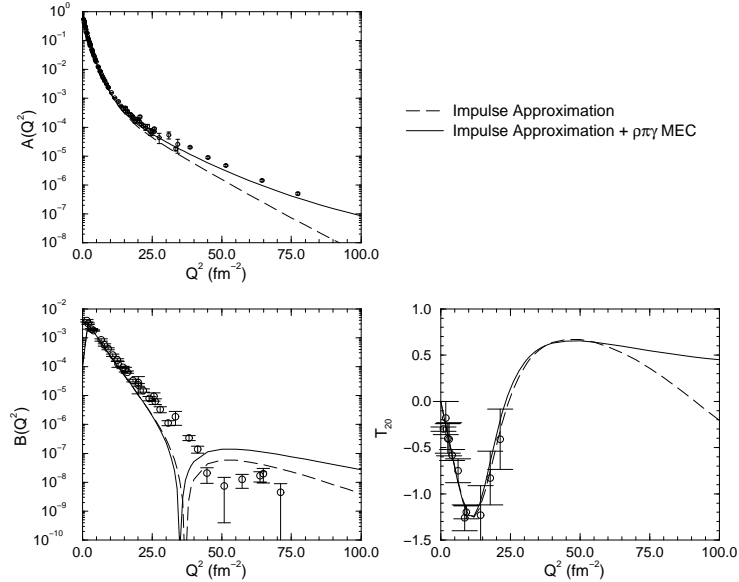 scaled 500}}

\caption{The form factors $A(Q^2)$ and $B(Q^2)$ together with the 
  tensor polarization for the deuteron. The long dashed line is an
  impulse approximation calculation with an instant interaction.  The
  solid line includes the effect of the $\rho \pi \gamma$
  MEC.}
\label{fig-RPG}
\end{figure}

This $\rho \pi \gamma$ MEC is a conserved current whose structure can
be found in Refs.~\cite{HT89,De94}. The couplings and form factors for
the meson-nucleon-nucleon vertices are all taken to be consistent with
those used in our one-boson-exchange interaction. Meanwhile, the $\rho
\pi \gamma$ coupling is set to the value $g_{\rho \pi \gamma}=0.56$,
and a vector meson dominance form factor is employed at the $\rho \pi
\gamma$ vertex: $F_{\rho \pi \gamma}(q)=1/(q^2 - m_\omega^2)$.  The
value of this MEC is added to the impulse contribution calculated
above and $A$, $B$, and $T_{20}$ are calculated. This is done with the
vertex function obtained from an instant interaction, and consequently
the electromagnetic current is exactly conserved. The results of this
calculation are displayed in Fig.~\ref{fig-RPG}. We see that at $Q^2$
of order 2 ${\rm GeV}^2$ the $\rho \pi \gamma$ MEC makes a significant
contribution to all three observables. However, far from improving the
agreement of the position of the minimum in the $B$ form factor with
the experimental data, this particular MEC moves the theoretical result
{\it away} from the data---as noted by Hummel and Tjon~\cite{HT89},
and seen within a simplified version of the formalism presented here
by Devine~\cite{De94}.  Thus, it would seem that some physics beyond
the impulse approximation other than the $\rho \pi \gamma$ MEC plays a
significant role in determining the position of the minimum in
$B(Q^2)$.

\section{Conclusion}

\label{sec-Section7}

A systematic theory of the electromagnetic interactions of
relativistic bound states is available in three dimensions. In this
formalism integrations are performed over the zeroth component of the
relative momentum of the two particles, leading to the construction of
``equal-time'' (ET) Green's functions.  If the formalism is to
incorporate the Z-graphs that are expected in a quantum field theory,
then the propagator must include terms coming from crossed Feynman
graphs.  Here we have displayed a three-dimensional propagator that
includes these effects correctly to leading order in $1/M$.

Given a suitable choice for the ET propagator, the electromagnetic and
interaction currents which should be used with it can be calculated.
If these are truncated in a fashion consistent with the truncation of
the $NN$ interaction in the hadronic field theory then the
Ward-Takahashi identities are maintained in the three-dimensional
theory.  A full accounting of the dynamical role played by
negative-energy states and of retardations in electromagnetic
interactions of the deuteron is thereby obtained.

Calculations have been performed for both the impulse approximation
and when the $\rho \pi \gamma$ MEC is included. In our MEC
calculations we use an instant approximation for the electromagnetic
current. This current satisfies current conservation when used with
deuteron vertex functions that are calculated with instant interactions.
We also have used this simpler current with vertex functions which 
are calculated with the retarded interactions obtained within the
ET formalism. 

Comparing impulse approximation calculations with and without
negative-energy states indicates that the role played by
negative-energy state components of the deuteron vertex function is
small. This corroborates the results of Hummel and Tjon and is in
contrast to those obtained in Ref.~\cite{vO95}. Because the ET
formalism incorporates the relevant Z-graphs in a preferable way, we
are confident that these Z-graphs really do play only a minor role in
calculations that are based upon standard boson-exchange models of the
$NN$ interaction.

The results for impulse approximation calculations of the
electromagnetic observables are relatively insensitive to the
distinction between a vertex calculated with retardations included and
one calculated in the instantaneous approximation. The results of both
calculations fall systematically below experimental data for the form
factors $A$ and $B$ for $Q$ of order 1 GeV.  This deficiency at higher
$Q$ suggests that mechanisms other than the impulse approximation
graph should be significant. Indeed, when the $\rho \pi \gamma$ MEC
graph is included in our calculation it somewhat remedies the result
for $A(Q^2)$. However, it fails to narrow the gap between our result
for $B(Q^2)$ and the existing experimental data. The significant gap
that remains between our theoretical result for $B(Q^2)$ and the data
indicates that it is an interesting observable in which to look for
physics of the deuteron other than the simple impulse mechanism or the
standard $\rho \pi \gamma$ MEC. Finally, the existing tensor
polarization data are reasonably well described. This is consistent
with previous analyses which have shown $T_{20}$ to be less sensitive
to non-impulse mechanisms.

\section*{Acknowledgments}
It is a pleasure to thank Steve Wallace for a fruitful and enjoyable
collaboration on this topic, and for his comments on this manuscript.
I am also very grateful to Neal Devine for giving us the original
version of the computer code to calculate these reactions, and to
Betsy Beise for useful information on the experimental situation.
Finally, I want to thank the organizers of this workshop for a
wonderful week of physics in Elba! This work was supported by the U.~S.
Department of Energy under grant no.  DE-FG02-93ER-40762.


\end{document}